\documentstyle[preprint,aps,epsfig]{revtex}

\begin{document}

\draft

\title{Correlations in Chaotic Eigenfunctions at Large Separation}

\author{Sanjay Hortikar \footnote{E--mail: \tt horti@physics.ucsb.edu}
        and 
        Mark Srednicki \footnote{E--mail: \tt mark@tpau.physics.ucsb.edu} }

\address{Department of Physics, University of California,
         Santa Barbara, CA 93106 }

\maketitle

\tighten

\begin{abstract}
An energy eigenfunction in a classically chaotic system is known to have 
spatial correlations which (in the limit of small $\hbar$) are governed by a 
microcanonical distribution in the classical phase space.  This result is 
valid, however, only over coordinate distances which are small compared to 
any relevant classical distance scales (such as the cyclotron radius for a 
charged particle in a magnetic field).  We derive a modified formula for the
correlation function in the regime of large separation.  This then permits a
complete description, over all length scales, of the statistical properties of
chaotic eigenfunctions in the $\hbar\to 0$ limit.  Applications to quantum dots
are briefly discussed.
\end{abstract}


In a hamiltonian system which exhibits classical chaos throughout 
the accessible phase space, the quantum energy eigenvalues and 
eigenfunctions are known to have certain universal properties 
in the limit of small $\hbar$ (which, in practice, is achieved
at sufficiently high energy) \cite{leshouces,casati}.
We will be interested in the energy eigenfunctions; these can be 
characterized as random variables with a gaussian probability distribution
of the form \cite{berry77,ogh,pei,paei,prig1,prig2,alhas,sred96,ss96}
\begin{equation}
P(\psi|E) \propto \exp
            \left[-{\beta\over2}\int d^f\! q_1\int d^f\! q_2\ 
                   \psi^*({\bf q}_2)K({\bf q}_2,{\bf q}_1|E)\psi({\bf q}_1)
            \right]\;,
\label{prob}
\end{equation}
where $f$ is the number of degrees of freedom.
If the system is time-reversal invariant,
the eigenfunctions are real and $\beta=1$; if it is not,
they are complex and $\beta=2$.
The measure corresponding to eq.~(\ref{prob}) is the standard one
of euclidean quantum field theory:
${\cal D}\psi = \prod_{\bf q} d\psi({\bf q})$ for $\beta=1$ and
${\cal D}\psi = \prod_{\bf q} d\mathop{\rm Re}\psi({\bf q}) \,
                              d\mathop{\rm Im}\psi({\bf q})$
for $\beta=2$.  $P(\psi|E){\cal D}\psi$ represents the probability that
the actual eigenfunction $\psi_\alpha({\bf q})$ for $E=E_\alpha$ 
(a particular energy eigenvalue) is between $\psi({\bf q})$ and
$\psi({\bf q}) + d\psi({\bf q})$ for all coordinates $\bf q$.
The kernel $K({\bf q}_2,{\bf q}_1|E)$
is the functional inverse of the two-point correlation function
\begin{equation}
C({\bf q}_2,{\bf q}_1|E) \equiv \int \psi({\bf q}_2)\psi^*({\bf q}_1) 
                                     P(\psi|E) {\cal D}\psi \; .
\label{auto}
\end{equation}
The explicit formula for $C({\bf q}_2,{\bf q}_1|E)$ 
which was originally suggested by Berry \cite{berry77}
assumes a microcanonical probability density in the classical phase space,
\begin{equation}
C({\bf q}_2,{\bf q}_1|E) = {1\over{\bar\rho}(E)}
                           \int {d^f\! p \over (2\pi\hbar)^f}\; 
                                e^{i{\bf p}\cdot({\bf q}_2-{\bf q}_1)/\hbar}
                                \delta(E-H_W({\bf p},{\bf \bar q})) \; ,
\label{berry}
\end{equation}
where ${\bf \bar q} \equiv {1\over2}({\bf q}_1+{\bf q}_2)$, 
$H_W({\bf p},{\bf q})$ is the classical hamiltonian
(more specifically, it is the Weyl symbol of the hamiltonian operator),
and ${\bar\rho}(E)$ is the semiclassical density of states,
\begin{equation}
{\bar\rho}(E) = \int {d^f\! p \, d^f\! q \over (2\pi\hbar)^f}\; 
                     \delta(E-H_W({\bf p},{\bf q})) \; .
\label{rhobar}
\end{equation}
However, eq.~(\ref{berry}) only applies when the separation
$|{\bf q}_2-{\bf q}_1|$ is sufficiently small \cite{berry77}.
Consider, for example, the case $H={\bf p}^2/2m + V({\bf q})$;
we would not expect the correlations in an eigenfunction at two
points ${\bf q}_1$ and ${\bf q}_2$ to depend only on the
value of the potential at ${\bf \bar q}$ if
$V({\bf \bar q})$ has a significantly different value
than either $V({\bf q}_1)$ or $V({\bf q}_2)$.  Our goal, then, is to
find the correct formula for $C({\bf q}_2,{\bf q}_1|E)$ when
$|{\bf q}_2-{\bf q}_1|$ is large.

One way to motivate eq.~(\ref{berry}) is to consider 
the energy Green's function \cite{gutz67,bm72}
\begin{equation}
G({\bf q}_2,{\bf q}_1|E) \equiv 
\sum_\alpha {\psi_\alpha({\bf q}_2)\psi_\alpha^*({\bf q}_1) 
             \over E-E_\alpha+i\epsilon} 
\label{green}
\end{equation}
where $\epsilon\to 0^+$.  We then have
\begin{equation}
\sum_\alpha \psi_\alpha({\bf q}_2)\psi_\alpha^*({\bf q}_1)\delta(E-E_\alpha)
= {1\over 2\pi i}\left[  G({\bf q}_1,{\bf q}_2|E)^* 
                       - G({\bf q}_2,{\bf q}_1|E)   
                 \right] \; ,
\label{imgreen}
\end{equation}
and the exact density of states is
\begin{eqnarray}
\rho(E) &\equiv& \sum_\alpha \delta(E-E_\alpha) 
\label{rho} \\
        &=& {1\over 2\pi i} \int d^f\! q \left[  G({\bf q},{\bf q}|E)^* 
                                               - G({\bf q},{\bf q}|E)   
                                         \right] \; .  
\label{rho2}
\end{eqnarray}
While the exact Green's function clearly has singularities whenever
$E=E_\alpha$, its leading approximation ${\overline G}({\bf q}_2,{\bf q}_1|E)$ 
in the small-$\hbar$ limit is a smooth function of its arguments.  
Given this, eqs.~(\ref{auto},\ref{imgreen},\ref{rho}) 
make it natural to expect that
\begin{equation}
C({\bf q}_2,{\bf q}_1|E) = {1\over 2\pi i {\bar\rho}(E)} 
                           \left[  {\overline G}({\bf q}_1,{\bf q}_2|E)^* 
                                 - {\overline G}({\bf q}_2,{\bf q}_1|E)   
                           \right] \; .
\label{cexp}
\end{equation}
This formula can be derived explicitly \cite{paei}
in the theory of disordered metals,
where a white-noise random potential is added to the hamiltonian.
In this case, eq.~(\ref{cexp}) holds with $\overline G$ standing for
the Green's function averaged over the random potential.  In the
limit that the strength of the potential is large (which corresponds
to the $\hbar\to 0$ limit for chaotic systems), eq.~(\ref{prob}) 
for the eigenfunction probability can also be derived 
explicitly \cite{pei,paei,prig1,prig2,sred96}.
Corrections for finite potential strength (that is, finite $\hbar$)
can also be computed \cite{blamir},
and are complimentary to the calculations done here.

Let us now briefly recall the construction of 
${\overline G}({\bf q}_2,{\bf q}_1|E)$
in the $\hbar\to 0$ limit \cite{gutz67,bm72}.
The energy Green's function is related to the propagator
$\langle {\bf q}_2|e^{-iHt/\hbar}|{\bf q}_1\rangle$ via
\begin{equation}
G({\bf q}_2,{\bf q}_1|E) = {1\over i\hbar}\int_0^\infty dt \;
                            e^{i(E+i\epsilon)t/\hbar} 
                            \langle {\bf q}_2|e^{-iHt/\hbar}|{\bf q}_1\rangle
                            \; .
\label{greenprop}
\end{equation}
The propagator can be written as
\begin{equation}
\langle {\bf q}_2|e^{-iHt/\hbar}|{\bf q}_1\rangle
=\int {d^f\!p\over(2\pi\hbar)^f} \; 
      e^{i{\bf p}\cdot({\bf q}_2-{\bf q}_1)/\hbar} 
      (e^{-iHt/\hbar})_W({\bf p},{\bf \bar q}) \;,
\label{prop}
\end{equation}
where again ${\bf \bar q} = {1\over2}({\bf q}_1+{\bf q}_2)$, and
$A_W({\bf p},{\bf \bar q})$ denotes the Weyl symbol of the operator $A$;
in fact, eq.~(\ref{prop}) is simply the Fourier transform of the definition
of the  Weyl symbol.  For sufficiently small times, 
\begin{equation}
(e^{-iHt/\hbar})_W({\bf p},{\bf q}) \simeq 
e^{-iH_W({\bf p},{\bf q})t/\hbar} \; .  
\label{weyl}
\end{equation}
Inserting eqs.~(\ref{prop},\ref{weyl}) into eq.~(\ref{greenprop}),
and performing the time integral, we get the desired approximation,
\begin{equation}
{\overline G}({\bf q}_2,{\bf q}_1|E) 
=\int {d^f\!p\over(2\pi\hbar)^f} \; 
      e^{i{\bf p}\cdot({\bf q}_2-{\bf q}_1)/\hbar} \;
      {1\over E-H_W({\bf p},{\bf \bar q})+i\epsilon} \;.
\label{greenclass}
\end{equation}
The derivation of this formula is flawed, however,
since we integrated over all positive times
even though eq.~(\ref{weyl}) is valid only for short times.
A more careful analysis
shows that eq.~(\ref{greenclass}) is valid
in the limit of small $\hbar$, provided that $|{\bf q}_2-{\bf q}_1|$
is small enough so that the shortest classical path ${\bf q}(t)$ connecting
${\bf q}_1$ to ${\bf q}_2$ with energy $E$ is well approximated by 
a linear function of time.  Note that this criterion is purely classical;
$|{\bf q}_2-{\bf q}_1|$ can be sufficiently small even if it is
large compared with the quantum wavelength $\hbar/p$, where
$p=|{\bf p}|$ is the magnitude of the classical momentum when
${\bf q}={\bf \bar q}$.  With this caveat, eq.~(\ref{greenclass}), 
when inserted into eq.~(\ref{cexp}), immediately yields eq.~(\ref{berry}).

As this derivation shows, however, eq.~(\ref{berry}) is not valid if 
$|{\bf q}_2-{\bf q}_1|$ is too large.
In this case, we must use a different semiclassical formula for 
$G({\bf q}_2,{\bf q}_1|E)$, making a stationary phase approximation both
in the Feynman path integral representation of the propagator, 
and in the time integral of eq.~(\ref{greenprop}).
The well-known result is \cite{gutz67}
\begin{equation}
{\overline G}({\bf q}_2,{\bf q}_1|E) = 
{1\over i\hbar(2\pi i\hbar)^{(f-1)/2}} \sum_{\rm paths}
         |D_{\rm p}|^{1/2}e^{iS_{\rm p}/\hbar-i\nu_{\rm p}\pi/2} \; .
\label{greengutz}
\end{equation}
Here the sum is over all classical paths connecting 
${\bf q}_1$ to ${\bf q}_2$ with energy $E$ and action
\begin{equation}
S_{\rm p} = \int_{{\bf q}_1}^{{\bf q}_2} {\bf p}\cdot d{\bf q} \; .
\label{s}
\end{equation}
The index $\nu_{\rm p}$ counts the number of classical focal points
along the path, and the determinant $D_{\rm p}$ of second derivatives
of $S_{\rm p}$ is given by
\begin{equation}
D_{\rm p} = \det\pmatrix{ {\partial^2 S_{\rm p} \over
                            \partial{\bf q}_2 \partial{\bf q}_1 } &
                           {\partial^2 S_{\rm p} \over
                            \partial E \partial{\bf q_1}        } \cr
\noalign{\medskip}
                           {\partial^2 S_{\rm p} \over
                            \partial{\bf q}_2 \partial E        } &
                           {\partial^2 S_{\rm p} \over
                            \partial E^2                        } \cr } \; .
\label{det}
\end{equation}
It is eq.~(\ref{greengutz}) which should be used in eq.~(\ref{cexp}) when
the shortest classical path from ${\bf q}_1$ to ${\bf q}_2$ is not
(approximately) a linear function of time.  

If the system is time-reversal invariant, then the paths from 
${\bf q}_2$ to ${\bf q}_1$ are have the same set of values of
$S_{\rm p}$, $\nu_{\rm p}$, and $D_{\rm p}$ as the paths from 
${\bf q}_1$ to ${\bf q}_2$.  In this case, we have
\begin{equation}
C({\bf q}_2,{\bf q}_1|E) = {2\over{\bar\rho}(E)(2\pi\hbar)^{(f+1)/2}}
                           \sum_{\rm paths} |D_{\rm p}|^{1/2}
                           \cos[S_{\rm p}/\hbar-(2\nu_{\rm p}+f-1)\pi/4] 
\label{cnew}
\end{equation}
instead of eq.~(\ref{berry}) when $|{\bf q}_2 - {\bf q}_1|$ is large.

To illustrate the differences between eq.~(\ref{cnew}) and
eq.~(\ref{berry}), let us examine a few special cases.  
First, we consider an $f$-dimensional billiard in which
the straight-line path from ${\bf q}_1$ to ${\bf q}_2$ is
not blocked.  In the interior of the billiard,
$H={\bf p}^2/2m$, and eq.~(\ref{berry}) yields \cite{berry77}
\begin{equation}
C({\bf q}_2,{\bf q}_1|E) = V^{-1}\Gamma(f/2)
                           {J_{(f-2)/2}(kL) \over
                            (kL/2)^{(f-2)/2}      } \; ,
\label{berryb}
\end{equation}
where $V$ is the $f$-dimensional volume of the billiard,
$\hbar k=(2mE)^{1/2}$, $L=|{\bf q}_2-{\bf q}_1|$ is the length
of the straight-line path, $\Gamma(x)$ is the Euler gamma function,
and $J_\nu(x)$ is an ordinary Bessel function.  
In the case at hand, eq.~(\ref{berryb}) is valid even for large $L$.
The reason is that the straight-line path is a linear function of time,
and this is the condition needed for the validity of eq.~(\ref{berry}).
Turning to eq.~(\ref{cnew}), we note that 
${\bar\rho}(E)=k^f V/(4\pi)^{f/2}\Gamma(f/2)E$, and that
the straight-line path has $S_{\rm p}/\hbar = kL$ and 
$|D_{\rm p}| = m^2(2mE)^{(f-3)/2}/L^{f-1}$.  
If ${\bf q}_1$ and ${\bf q}_2$ are both far from any of
the billiard's walls, the straight-line path makes the dominant
contribution, and we find
\begin{equation}
C({\bf q}_2,{\bf q}_1|E) = V^{-1}\Gamma(f/2)
                           {\cos[kL
                                   -(2\nu_{\rm p}+f-1)\pi/4] \over
                            \pi^{1/2} (kL/2)^{(f-1)/2} } 
\label{newb}
\end{equation}
with $\nu_{\rm p}=0$.  Eq.~(\ref{newb})
is equivalent to eq.~(\ref{berryb}) when $kL$ is large, since
in this regime the asymptotic form of the Bessel function can be invoked.
Thus, in the present case, both eq.~(\ref{berry}) and eq.~(\ref{cnew})
are valid for large $L$, and eq.~(\ref{berry}) is valid for small $L$ 
(less than a quantum wavelength, $2\pi/k$) as well.

On the other hand, if we consider a billiard in which the straight-line
path from ${\bf q}_1$ to ${\bf q}_2$ is blocked by an obstacle, 
such as in fig.~(1), eq.~(\ref{berryb}) is
not correct, and we must use eq.~(\ref{cnew}).  The shortest classical
path connecting ${\bf q}_1$ to ${\bf q}_2$
makes the dominant contribution; this is given by eq.~(\ref{newb}),
provided we take $L$ to be the length of the path,
and set $\nu_{\rm p}$ equal to twice the number of bounces.
In most realistic cases of this type, there will be many other paths
with more bounces that are not very much longer;
these will all contribute to $C({\bf q}_2,{\bf q}_1|E)$ as well.
However, if we attempt to verify eq.~(\ref{newb}) numerically 
for a particular eigenfunction of a particular billiard, it is necessary
to average over a range of ${\bf q}_1$ and ${\bf q}_2$ in order to reduce
the variance in $C({\bf q}_2,{\bf q}_1|E)$ which is expected from the 
probability distribution (\ref{prob}) \cite{berry77,ogh,ss96}.
If this averaging is carried out with the length $L$ of the shortest
classical path held fixed, the other contributing paths will in general
have lengths that vary.  If this variation is large on the scale of
the quantum wavelength $2\pi/k$, then the net contribution of all these
other paths to the averaged $C({\bf q}_2,{\bf q}_1|E)$ should be small,
rendering eq.~(\ref{newb}) a valid formula for the averaged correlation
function.

As another example we consider a billiard with the addition of an 
isotropic harmonic potential $V({\bf q})={1\over2}m\omega^2{\bf q}^2$.
This case is of some physical interest for $f=2$; wave-function correlations
in quantum dots have been studied 
assuming that the dot is well-modeled by a two-dimensional
billiard, but in fact there is also a smooth confining 
potential, often approximated as harmonic.  For simplicity, we
give only the leading corrections in the limit of a weak potential.
The shortest classical path (assuming it is not blocked) has action
\begin{equation}
S_{\rm p}/\hbar = kL\left[1+{L^2-3d^2 \over  12 R^2} 
                             +\ldots\right] \; ,
\label{sp} 
\end{equation}
where $\hbar k=(2mE)^{1/2}$ and $L=|{\bf q}_2 -{\bf q}_1|$ as before,
and we have introduced
$d^2 \equiv {\bf q}_1^2 + {\bf q}_2^2$ and
$R \equiv (2E/m\omega^2)^{1/2}$; $R$ is the maximum distance from the origin 
which can be reached with energy $E$.  Also, we find
\begin{equation}
|D_{\rm p}| = |D_{\rm p}|_{\omega=0}
              \left[1+{(f-1)L^2-(f-3)d^2 \over 4 R^2} 
                             +\ldots\right] \; .
\label{dp}
\end{equation}
In all but two dimensions, the correction to $|D_{\rm p}|$ is
dominated by the correction to $\bar\rho(E)$, which is $O(V^{2/f}/R^2)$,
where $V^{1/f}$ is the linear size of the billiard.
However, when $f=2$, $\bar\rho(E)$ is independent of $E$,
and it is not changed by the presence of a weak potential.
In this case, eq.~(\ref{dp}) represents the dominant correction
to the amplitude of $C({\bf q}_2,{\bf q}_1|E)$.

As a final example with importance for quantum dots,
we consider a particle with charge $e$ in a two-dimensional billiard with
a uniform, perpendicular magnetic field $\bf B$.  In the billiard
interior we have $H=({\bf p}-e{\bf A})^2/2m$, and we
will work in the gauge in which the vector potential is
${\bf A}={1\over2}{\bf B}\times{\bf q}$.  This system
is not time-reversal invariant, and so we must use eqs.~(\ref{cexp})
and (\ref{greengutz}) rather than eq.~(\ref{cnew}).  We again consider
points ${\bf q}_1$ and ${\bf q}_2$ which are far from the billiard's walls.
The shortest classical path is then a circular arc with length $\ell$,
related to the separation $L=|{\bf q}_2 -{\bf q}_1|$ and classical cyclotron
radius $R \equiv (2mE)^{1/2}/|eB|$ via $\ell=2R\sin^{-1}(L/2R)$.  The action
for this path can be divided into a geometric part and a gauge-dependent part,
$S_{\rm p} = S_{\rm geom} + S_{\rm gauge}$.  
The geometric part is
\begin{eqnarray}
S_{\rm geom}  &=& \hbar k \left(\ell-{{\cal A}\over R}\right) 
\nonumber \\
\noalign{\medskip}
              &=& \hbar kL\left(1-{L^2\over 24 R^2} +\ldots\right) \; ,
\label{sgeo} 
\end{eqnarray}
where again $\hbar k=(2mE)^{1/2}$,
and ${\cal A} = {1\over2}R \ell - {1\over2} R^2 \sin(\ell/R)$ 
is the area enclosed by the circular arc and the straight line connecting
${\bf q}_1$ to ${\bf q}_2$.  
The gauge-dependent part is energy independent, and changes sign when 
${\bf q}_1$ and ${\bf q}_2$ are exchanged.  For our gauge choice,
\begin{equation}
S_{\rm gauge} = {\textstyle{1\over 2}} 
                  e {\bf B}\!\cdot\!({\bf q}_1\times{\bf q}_2) \; .
\label{sgauge}
\end{equation}
If we make a gauge transformation 
\begin{equation}
{\bf A}({\bf q}) \to {\bf A}({\bf q}) + \nabla\Phi({\bf q}) \; ,
\label{trans1}
\end{equation}
where $\Phi({\bf q})$ is any smooth function, then
\begin{equation}
S_{\rm gauge} \to S_{\rm gauge} + e\Phi({\bf q}_2) - e\Phi({\bf q}_1) \; .
\label{trans2}
\end{equation}
The determinant $|D_{\rm p}|$, on the other hand, is gauge invariant,
\begin{equation}
|D_{\rm p}| = {m^2 \over \hbar kL}\left(1-{L^2\over 4 R^2}\right)^{-1/2} \; .
\label{dpb}
\end{equation}
Again there is no correction to $\bar\rho(E)$ in two dimensions.
Keeping only the contribution of this path, we find from 
eqs.~(\ref{cexp}) and (\ref{greengutz}) that
\begin{equation}
C({\bf q}_2,{\bf q}_1|E) = V^{-1}\exp(iS_{\rm gauge}/\hbar)
                           {\cos(S_{\rm geom}/\hbar -\pi/4) \over
                           (\pi kL/2)^{1/2}
                           (1-L^2/4 R^2)^{1/4} } \; ,
\label{cb}
\end{equation}
where $V$ is the area of the billiard.  Under the gauge transformation
(\ref{trans1}), eq.~(\ref{trans2}) implies
\begin{equation}
C({\bf q}_2,{\bf q}_1|E) \to e^{+ie[\Phi({\bf q}_2)-\Phi({\bf q}_1)]/\hbar}
                             C({\bf q}_2,{\bf q}_1|E) \; .
\label{ctr}
\end{equation}
That this is correct can be seen by recalling that a wave function
$\psi({\bf q})$ transforms as 
\begin{equation}
\psi({\bf q)} \to e^{+ie\Phi({\bf q})/\hbar}\psi({\bf q)}
\label{psitr}
\end{equation}
under (\ref{trans1}), and that $C({\bf q}_2,{\bf q}_1|E_\alpha)$
is the expected value of $\psi_\alpha({\bf q}_2)\psi_\alpha^*({\bf q}_1)$.
On the other hand, eq.~(\ref{berry}) implies that 
\begin{equation}
C({\bf q}_2,{\bf q}_1|E) \to e^{+ie({\bf q}_2-{\bf q}_1)\cdot
                                   \nabla\Phi({\bf \bar q})/\hbar}
                             C({\bf q}_2,{\bf q}_1|E) \; ,
\label{cbtr}
\end{equation}
which again illustrates the fact that eq.~(\ref{berry}) is valid only when
$|{\bf q}_2-{\bf q}_1|$ is sufficiently small.

Finally, we note that our expression for
$C({\bf q}_2,{\bf q}_1|E)$ is needed to resolve
a discrepancy between two different formulas in the literature
for the expected values of off-diagonal matrix elements 
(in the energy-eigenstate basis) of simple, $\hbar$-independent
operators in classically chaotic systems.
This will be the subject of a separate paper.

\begin{acknowledgments}

This work was supported in part by NSF Grant PHY--97--22022.

\end{acknowledgments}

\begin{figure}
\centerline{ \epsfxsize=200pt  \epsfysize=200pt  \epsfbox{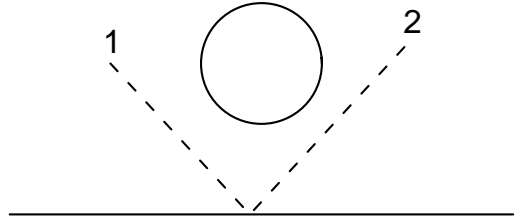} }
\caption{
A portion of a Sinai billiard in which the circular scatterer
is off center; the path shown is the shortest classical path
between points one and two.
}
\label{fig1}
\end{figure}


\begin{references}

\bibitem{leshouces} {\em Les Houches LII, Chaos and Quantum Physics}, 
ed. M.-J. Giannoni, A. Voros, and J. Zinn-Justin 
(North--Holland, Amsterdam, 1991).

\bibitem{casati} {\em Quantum Chaos},
ed. G. Casati and B. Chirikov
(Cambridge Univ. Press, New York, 1995).

\bibitem{berry77} M. V. Berry, J. Phys. A {\bf 10}, 2083 (1977).

\bibitem{ogh} P. O'Connor, J. Gehlen, and E. J. Heller,
Phys. Rev. Lett. {\bf 58}, 1296 (1987).

\bibitem{pei} V. N. Prigodin, K. B. Efetov, and S. Idia
Phys. Rev. Lett. {\bf 71}, 1230 (1993).

\bibitem{paei} V. N. Prigodin, B. L. Alshuler, K. B. Efetov, and S. Idia
Phys. Rev. Lett. {\bf 72}, 546 (1994).

\bibitem{prig1} V. N. Prigodin,
Phys. Rev. Lett. {\bf 74}, 1566 (1995).

\bibitem{prig2} V. N. Prigodin, N. Taniguchi, A. Kudrolli, V. Kidambi,
and S. Sridhar, Phys. Rev. Lett. {\bf 75}, 2392 (1995).
(cond-mat/9504089)

\bibitem{alhas} Y. Alhassid and C. H. Lewenkopf,
Phys. Rev. Lett. {\bf75}, 3922 (1995).
(cond-mat/9510023)

\bibitem{sred96} M. Srednicki, Phys. Rev. E {\bf 54}, 954 (1996).
(cond-mat/9512115)

\bibitem{ss96} M. Srednicki and F. Stiernelof,
J. Phys. A {\bf 29}, 5817 (1996).
(chao-dyn/9603012)

\bibitem{gutz67} M. C. Gutzwiller, J. Math. Phys. {\bf 8}, 1979 (1867).

\bibitem{bm72} M. V. Berry and K. E. Mount,
Rep. Prog. Phys. {\bf 35}, 315 (1972).

\bibitem{blamir} Ya. M. Blanter and A. D. Mirlin,
Phys. Rev. E {\bf 55}, 6514 (1997).
(cond-mat/9604139)

\end{references}
\end{document}